\documentclass[onecollarge,natbib]{svjour2}
\bibpunct{[}{]}{;}{n}{}{,} 
\smartqed  
\usepackage{graphicx}
\usepackage{graphics}
\def\be{\begin{equation}}
\def\ee{\end{equation}}
\def\bea{\begin{eqnarray}}
\def\eea{\end{eqnarray}}
\def\bear{\begin{array}}
\def\eear{\end{array}}
\newcommand{\MSbar}{\overline{\rm MS}}

\newcommand{\ta}{{\widetilde a}}
\newcommand{\td}{{\widetilde d}}

\newcommand{\tk}{{\widetilde k}}

\newcommand{\tal}{{\widetilde \alpha}}

\begin{document}

\title{Evaluations of low-energy physical quantities in QCD with IR freezing of the coupling}

\titlerunning{Evaluations in QCD with IR freezing}        

\author{Gorazd Cveti\v{c}}
\institute{Dept.~of Physics, Universidad T\'ecnica
Federico Santa Mar\'{\i}a (UTFSM), Valpara\'{\i}so, Chile}


\institute{Dept.~of Physics, Universidad T\'ecnica
Federico Santa Mar\'{\i}a (UTFSM), Valpara\'{\i}so, Chile}

\date{Based on the presentation at the Satellite Meeting 
of Lightcone 2013+, Skiathos, Greece, May 24-25, 2013. Preprint USM-TH-317}

\maketitle

\begin{abstract}
The ${\overline{\rm MS}}$-like schemes in QCD have in general the running coupling which
contains Landau singularities, i.e., singularities 
outside the timelike semi-axis, at low squared momenta. 
As a consequence, evaluation of the spacelike quantities, such as
current correlators, in terms of (powers of) such a coupling then results in
quantities which contradict the basic principles of Quantum Field Theories.
On the other hand, in those QCD frameworks where the running coupling remains 
finite at low squared momenta (IR freezing), the coupling usually does not 
have Landau singularities in the complex plane of the squared momenta. 
I argue that in such QCD frameworks the spacelike quantities should not be evaluated 
as a power series, but rather as a series in derivatives of the coupling with respect 
to the logarithm of the squared momenta. Such series show considerably better 
convergence properties. Moreover, Pad\'e-related resummations of such 
logarithmic derivative series give convergent series, thus eliminating 
the practical problem of series divergence due to renormalons.
\keywords{low-energy QCD, IR freezing, logarithmic derivatives, Pad\'e-related resummation}
\end{abstract}

\section{Introduction}
\label{sec:intro}
One of the main challenges of the contemporary particle physics is to understand
and adequately describe QCD at low scales $\stackrel{<}{\sim} 1$ GeV.
The usual perturbative (pQCD) approach in $\MSbar$-like schemes leads to the
running coupling  $a(Q^2)$ ($\equiv \alpha_s^{\rm (pt)}(Q^2)/\pi$) which has
singularities in the regime outside the negative axis 
in the complex $Q^2$-plane (where $q^2 \equiv - Q^2$
is the usual squared momentum transfer). Such singularities are not present
in the spacelike renormalization scale invariant quantities $d(Q^2)$, as a
consequence of the basic principles of quantum field theories \cite{BS}
such as locality, unitarity and microcausality. 
If such quantities are to be evaluated as functions
of the running coupling $a(\kappa Q^2)$ (with $\kappa \sim 1$), then the coupling
$a$ should not have such (Landau) singularities. Thus $a(Q^2)$ should be an
analytic (holomorphic) function of $Q^2$ in the entire complex plane, with the
exception of the negative semiaxis $Q^2 < - M^2_{\rm thr}$ (where $ M^2_{\rm thr} \sim
10^{-1} \ {\rm GeV}^2$). 

Such a behavior of $a(Q^2)$ is indirectly supported by calculations
using functional methods 
\cite{DSE1a,DSE1b,DSE2a,DSE2b,FRGa,FRGb,FRGc,STQ} 
and  lattice calculations \cite{lattice1a,lattice1b,lattice2a,lattice2b}. 
Most of these works suggest that
the running coupling has a finite limit when $Q^2 \to 0$, i.e., IR freezing
(IR fixed point). IR freezing is obtained also in models with  AdS/CFT
correspondence modified by a dilaton backgound \cite{AdS,Lyubovitskij}.

In the works \cite{ShS,MSa,MSb,Sh1Sh2,KS,KuMagr,BMS1,BMS2,BMS3,Webber,1danQCD,2danQCD}
the mentioned type of analyticity was imposed on the QCD coupling $a(Q^2)$, within
various scenarios, and as a result the obtained holomorphic coupling 
turned out to be IR finite (for reviews, see \cite{Bakulev,Stefanis}). 
Infrared finiteness of the coupling and its analyticity,
however, do not necessarily always go together. For example,
a model with holomorphic coupling which is infinite in the limit $Q^2 \to 0$ was
constructed and used in  Refs.~\cite{Nesterenko1a,Nesterenko1b}. 
The opposite example is that of
Ref.~\cite{StevensonMatt} where the coupling is finite in the limit $Q^2 \to 0$ but
has (Landau) singularities within the complex $Q^2$ plane outside the real axis
(see the comments about this coupling in Ref.~\cite{panQCD}).

Yet another question is whether a purely perturbative coupling $a(Q^2)$, in the
$\MSbar$-like schemes, can have a holomorphic (and IR finite) coupling $a(Q^2)$
[$\MSbar$-like schemes are speficied later on in the comments after Eq.~(\ref{pRGE})].
In Ref.~\cite{panQCD} it was shown that such schemes are difficult to obtain,
and appear to lead to sudden jumps in the values of the coefficients $\beta_j$ of
the beta function when $j$ increases. Yet there exist QCD models with holomorphic and
IR finite coupling $a(Q^2)$ which practically merge with the 
underlying pQCD couplings $a_{\rm pt}(Q^2)$
in the $\MSbar$-like schemes at higher $|Q^2|$, i.e., $a(Q^2)-a_{\rm pt}(Q^2) \sim (Q^2/\Lambda^2)^N$
at  $Q^2 > \Lambda^2$ (where $\Lambda^2 \sim 10^{-1} \ {\rm GeV}^2$) with $N=4,5$,
Refs.~\cite{Alekseev,Webber,1danQCD,2danQCD}. In particular, the analytic model
\cite{2danQCD} has $N=5$, and reproduces the experimental value of
the $\tau$ lepton semihadronic (strangeless) decay ratio $r_{\tau}$,
the latter quantity being 
one of the few well measured low-momenta QCD quantities at present.

Here I will present three frameworks with IR finite coupling $a(Q^2)$ which, in addition,
is holomorphic in the $Q^2$ complex plane (with the exception of a semiaxis $Q^2 < - M^2_{\rm thr}$).
I will argue that the renormalization scale invariant spacelike quantities 
(such as spacelike observables) $d(Q^2)$ at low $|Q^2| \sim 1\ {\rm GeV}^2$
should not be evaluated, as usually assumed, 
as a series in powers $a(\kappa Q^2)^{n+1}$, but rather as a series in logarithmic derivatives
$\ta_{n+1}(\kappa Q^2) \propto (\partial/\partial \ln Q^2)^{n} a(\kappa Q^2)$. 
This, because the renormalization scale dependence of the (truncated) power series 
grows out of control and the series shows strong divergence (compounded by the renormalon problem)
when the number of terms increases. Further, I consider a 
resummation method of Refs.~\cite{BGApQCD1a,BGApQCD1b,GCRK,anOPE},
which is based on the mentioned truncated series in logarithmic derivatives 
and is a generalization of the diagonal Pad\'e method. I show that this
generalized Pad\'e method in the frameworks with IR finite coupling gives results which
are renormalization scale invariant and converge very well when the number of terms
in the initial truncated series increases. Numerical evidence is presented
for the large-$\beta_0$ Adler function, which is a spacelike renormalization scale invariant
quantity whose expansion is known to all orders. Finally, I argue that the obtained
conclusions are applicable also to the timelike observables, because they can be represented
as integral transformations of the aforementioned spacelike quantities.
A more detailed consideration of these topics has been presented in 
Ref.~\cite{GC}.

\section{Three scenarios with IR finite (and holomorphic) coupling}
\label{sec:3scen}

To fix the notations, I start here with the (truncated) perturbative RGE
\begin{eqnarray}
\frac{\partial a_{\rm pt}(Q^2; {\beta_2}, \ldots)}
{\partial \ln Q^2} 
& = &
- \sum_{j=0}^{n-1} \beta_{j} \: 
a_{\rm pt}(Q^2; {\beta_2}, \ldots)^{j+2} = 
- \beta_0 a_{\rm pt}^2 (1 + c_1 a_{\rm pt} + c_2 a_{\rm pt}^2 + \ldots) \ ,
\label{pRGE}
\end{eqnarray}
where $a_{\rm pt} \equiv \alpha_s/\pi = g_s/(4 \pi^2)$, 
the first two beta coefficients are universal 
[$\beta_0= (1/4)(11-2 N_f/3)$, $\beta_1=(1/16)(102-38 N_f/3)$],
and the other coefficients $\beta_k \equiv c_j \beta_0$ ($k \geq 2$) characterize
the perturbative renormalization scheme. The renormalization schemes are
called $\MSbar$-like if the coefficients $\beta_j$ depend on the (quark) mass via
the number of effective quark flavors $N_f$ and are polynomials of $N_f$ of order
$j$ for $j \geq 2$.
For the $\Lambda$-scale convention
($\Lambda$ ``scheme'') I take $\Lambda_{\MSbar}$. 

\subsection{Coupling with dynamical gluon mass}
\label{sec:dgm}

A representative case of QCD coupling $a(Q^2)$ with finite $Q^2 \to 0$ limit is the
case with effective (dynamical) gluon mass $m$, 
Refs.~\cite{Simonov1,Simonov2,BKS,KKSh}
\be
a^{(m)}(Q^2) = a_{\rm pt}(Q^2+m^2) \ ,
\label{mcase}
\ee
where I take $m=0.8$ GeV, $N_f=3$, and $a_{\rm pt}$ as the usual pQCD coupling in the
$c_2=c_3=\ldots=0$ renormalization scheme which allows exact solution in terms of the
Lambert function, Refs.~\cite{Gardi:1998qr,Magr} 
(see also Ref.~\cite{GarKat})
\bea
a_{\rm pt}(\kappa Q^2) = - \frac{1}{c_1} \frac{1}{\left[
1 + W_{\mp 1}(z) \right]} \ .
\label{aptexact}
\eea
Here, $Q^2=|Q^2| \exp(i \phi)$; $W_{-1}$ and $W_{+1}$
are the branches of the Lambert function
for $0 \leq \phi < + \pi$ and $- \pi < \phi < 0$, 
respectively, and $z$ is defined as
\be
z =  - \frac{1}{c_1 e} 
\left( \frac{\kappa |Q^2|}{\Lambda_{\rm L.}^2} \right)^{-\beta_0/c_1} 
\exp \left( - i {\beta_0}\phi/c_1 \right) \ ,
\label{zexpr}
\ee 
where $\Lambda_{\rm L.}$ is the Lambert QCD scale. At $N_f=3$ we have 
$\Lambda_{\rm L.} = \Lambda_{\MSbar}/0.72882$. I use $\Lambda_{\rm L.}=0.487$ GeV, 
thus $\Lambda_{\MSbar}=0.355$ GeV. This gives at $\mu^2=m_{\tau}^2$ 
the value $a^{(m)}(m_{\tau}^2) = 0.293/\pi$.

\subsection{(Fractional) Analytic Perturbation Theory (F)APT}
\label{sec:fapt}

This is the model developed in 
\cite{ShS,MSa,MSb,Sh1Sh2,KS,KuMagr,BMS1,BMS2,BMS3}.
The analogs $a^{\rm (FAPT)}_{\nu}(Q^2)$ of the power $a_{\rm pt}(Q^2)^{\nu}$
(where $\nu$ can be noninteger; and $a_{\rm pt}$ is in a $\MSbar$-like 
renormalization scheme) 
are obtained by ``minimally'' analytizing
the pQCD expression $a_{\rm pt}(Q^2)^{\nu}$. This means that the
cuts of  $a_{\rm pt}(Q^2)^{\nu}$ on the negative axis $Q^2 \equiv -\sigma < 0$ are
kept unchanged, but the Landau singularities (cuts and poles) on the positive
$Q^2$ axis are eliminated. This leads via Cauchy theorem to the following
dispersive expression for $a_{\nu}$:
\begin{equation}
a^{\rm {(FAPT)}}_{\nu}(Q^2) = \frac{1}{\pi} \int_{\sigma= 0}^{\infty}
\frac{d \sigma {\rho^{\rm {(pt)}}_{\nu}}(\sigma) }{(\sigma + Q^2)} \qquad
\left[ \not= a^{\rm {(FAPT)}}(Q^2)^{\nu} \right] \ ,
\label{MAAnudisp} 
\end{equation}
where ${\rho^{\rm {(pt)}}_{\nu}}(\sigma) = {\rm Im} a_{\rm pt}(-\sigma - i \epsilon)^{\nu}$
is the discontinuity function on the cut. 
At one-loop level $a_{\nu}^{\rm (FAPT)}$ has an explicit expression
and was constructed and used in Ref.~\cite{BMS1}
\be
a_{\nu}(Q^2)^{\rm (FAPT, 1-\ell.)} = \frac{1}{\beta_0^{\nu}}
\left(  \frac{1}{\ln^{\nu}(z)} -
\frac{ {\rm Li}_{-\nu+1}(1/z)}{\Gamma(\nu)} \right) \ ,
\label{MAAnu1l}
\ee
where $z \equiv Q^2/\Lambda^2$ and 
${\rm Li}_{-\nu+1}(z)$ is the polylogarithm function of order $-\nu+1$.
FAPT expressions for higher loops can be obtained via expansions of the
one-loop result \cite{BMS2,BMS3}. A review of FAPT is given in 
Refs.~\cite{Bakulev,Stefanis}. Mathematical packages 
for numerical calculation are given in Refs.~\cite{BK1a,BK1b,BK2}. 
I will use for the underlying renormalizations scheme $c_2=c_3=\cdots = 0$,
and for the number of active quark flavors $N_f=3$.
The (F)APT scale is fixed at  ${\Lambda}_{\rm L.}({\rm (F)APT}) = 0.572$ GeV, 
giving the value $a^{\rm (FAPT)}(m_{\tau}^2)=0.295/\pi$. 

\subsection{Analytic model with two deltas  (2$\delta$anQCD)}
\label{sec:2danQCD}

This model also has holomorphic $a(Q^2)$, and is based on the general dispersive relation 
for such couplings,
\begin{equation}
a(Q^2) = \frac{1}{\pi} \int_{\sigma= 0}^{\infty}
\frac{d \sigma {\rho}(\sigma) }{(\sigma + Q^2)} \ ,
\label{A1disp}
\end{equation}
where  ${\rho}$ is the discontinuity function of $a$:
${\rho}(\sigma) = {\rm Im} a(-\sigma - i \epsilon)$.
In Ref.~\cite{2danQCD} this discontinuity function
was approximated at high scales $\sigma \geq M_0^2$ (${\sim} 1 \ {\rm GeV}^2$)
by its pQCD analog $\rho^{\rm (pt)}(\sigma)={\rm Im} a_{\rm pt}(-\sigma - i \epsilon)$.
In the unknown low-scale regime, $0 < \sigma < M_0^2$
it was approximated by two delta functions
\bea
\rho(\sigma)^{(2 \delta)}(\sigma) & = & \pi F_1^2 \delta(\sigma - M_1^2)
+ \pi F_2^2 \delta(\sigma - M_2^2) + \Theta(\sigma-M_0^2) \rho^{\rm (pt)}(\sigma) \ .
\label{rho2d}
\eea
This gives via the dispersion relation (\ref{A1disp}) the following
coupling:
\bea
a^{(2 \delta)}(Q^2) &=&  \frac{F_1^2}{Q^2 + M_1^2} +
\frac{F_2^2}{Q^2 + M_2^2}
+ \frac{1}{\pi} \int_{M_0^2}^{\infty} d \sigma \; 
\frac{\rho^{\rm (pt)}(\sigma)}{(Q^2+\sigma)} \ .
\label{2dA1}
\eea
The parameters $F_j$ and $M_j$ ($j=1,2$) appearing in
the delta functions, and the pQCD-onset scale $M_0$, were
adjusted so that the correct value of the semihadronic tau decay
ratio $r_{\tau} \approx 0.20$ ($V+A$ channel) was reproduced and that
the difference from the underlying pQCD coupling at high $|Q^2| > \Lambda^2$
is as strongly suppressed as possible
\bea
a^{(2 \delta)}(Q^2) - a_{\rm pt}(Q^2) & \sim & 
(\Lambda^2/Q^2)^5 \ .
\label{dif2d} 
\eea
The renormalization scheme parameter value $c_2=-4.76$ of the underlying 
$N_f=3$ coupling $a_{\rm pt}$
was chosen in such a way that $M_0$ and the value
of $a^{(2 \delta)}(Q^2=0)$ were reasonable, i.e., not too high:
$M_0=1.25$ GeV and $a(0) \approx 0.78$ ($c_2$ can be varied between
$-5.7$ and $-2.1$, see Table I in Ref.~\cite{CAGC}).
In addition, it was convenient to choose $c_j = c_2^{j-1}/c_1^{j-2}$  ($j=3,4,\ldots$),
because then the exact solution of the underlying pQCD coupling is also known
in terms of the Lambert function (Refs.~\cite{Gardi:1998qr,KuMagr}, cf.~also
Ref.~\cite{CveKon}). I refer for more details on the model to Ref.~\cite{2danQCD}.
The input values of the model are the central ones used in Ref.~\cite{2danQCD}
(among them: $c_2=-4.76$, $\Lambda_{\rm L.} = 0.260$ GeV)
and give the value $a^{\rm (2 \delta)}(m_{\tau}^2)=0.291/\pi$.

\section{Series in powers and logarithmic derivatives}
\label{sec:powlder}


A spacelike QCD quantity $d(Q^2)$ with renormalization scale invariance, 
such as the derivative of a current correlator, is usually evaluated
in $\MSbar$-like schemes as a truncated power series
\be
d(Q^2; \kappa)_{\rm pt}^{[N]} = a_{\rm pt}(\kappa Q^2) + 
\sum_{j=1}^{N - 1} d_j(\kappa) \; a_{\rm pt}(\kappa Q^2)^{j+1} \ ,
\label{Dpttr}
\ee
where $\mu^2 \equiv \kappa Q^2$ is the renormalization scale
($\kappa \sim 1$), and usually $N=3$ or $N=4$.
Due to truncation, there appears the dependence on the renormalization
scale parameter $\kappa$
\be
\frac{ \partial d_{\rm pt}^{[N]}}{\partial \ln \kappa}
= K_N a_{\rm pt}(\kappa Q^2)^{N+1} +  K_{N+1} a_{\rm pt}(\kappa Q^2)^{N+2} + \cdots 
\; \sim a^{N+1} \ .
\label{RSdepDpt}
\ee 
This dependence may be quite large at low $Q^2$ and large $N$, one reason being 
the increase of the coefficients $K_{N+k}$ ($\sim d_{N+k-1}$) when $N+k$ increases
(due to renormalon growth); the other reason is the increase of
$a_{\rm pt}(\kappa Q^2)^{N+k+1}$ when $N+k$ increases because $a_{\rm pt}(\kappa Q^2)$ 
is large due to vicinity of the Landau singularities (at low positive $Q^2$). 
These two reasons also result in a very strongly divergent behavior of the
truncated power series (\ref{Dpttr}) when the number of terms $N$ increases
and $|Q|$ is low.

However, the power series can be reorganized in a series of
logarithmic derivatives \cite{CV1,CV2,panQCD}
\be
{\ta}_{{\rm pt},n}(Q^2)
\equiv \frac{(-1)^{n-1}}{\beta_0^{n-1} (n-1)!}
\left( \frac{ \partial} {\partial \ln Q^2} \right)^{n-1}  
a_{\rm pt}(Q^2) \ ,
\qquad (n=1,2,\ldots) \ .
\label{tan}
\ee
It can be shown by the RGE (\ref{pRGE}) that 
\be
{\ta}_{{\rm pt},n}(Q^2) = a_{\rm pt}(Q^2)^{n} 
+ \sum_{m \geq 1} k_m(n) a_{\rm pt}(Q^2)^{n+m} \ ,
\label{tanvsan}
\ee
where $k_m(n)$ depend on the coefficients $c_j$ of the RGE (\ref{pRGE}). These
relations can be inverted
\be
{a}_{{\rm pt}}(Q^2)^n = {\ta}_{{\rm pt},n}(Q^2) 
+ \sum_{m \geq 1} {\tk}_m(n) {\ta}_{{\rm pt},n+m}(Q^2) \ .
\label{tanan}
\ee
Inserting these expressions in the truncated power series (\ref{Dpttr}) results
in the reorganized truncated series (mpt) in the logarithmic derivatives
\be
d(Q^2; \kappa)_{\rm mpt}^{[N]} = a_{\rm pt}(\kappa Q^2) + 
\sum_{j=1}^{N - 1} \td_j(\kappa) \; \ta_{{\rm pt},j+1}(\kappa Q^2) \ .
\label{Dmpttr}
\ee
The two series (\ref{Dpttr}) and (\ref{Dmpttr}) differ in terms 
$\sim a_{\rm pt}^{N+1} \sim \ta_{{\rm pt},N+1}$ due to truncation.
Further, the renormalization scale dependence has now a different,
more simple, expression than in the case of the truncated power series (\ref{RSdepDpt})
\be
\frac{ \partial d_{\rm mpt}^{[N]}}{\partial \ln \kappa}
= -\beta_0 N \td_{N-1}(\kappa) \ta_{{\rm pt},N+1}(\kappa Q^2) \ .
\label{RSdepDmpt}
\ee 

In pQCD with $\MSbar$-like scheme, the two approaches of evaluation
give comparable results, even at low $|Q|$, as demonstrated in Ref.~\cite{GCCVMar}. 
However, in QCD with coupling $a(Q^2)$ finite in the IR regime, 
 at low $|Q|$ the method (\ref{Dmpttr}) with logarithmic derivatives
is significantly better than (\ref{Dpttr}) and, in fact, 
is the correct one, as argued in Refs.~\cite{CV1,CV2} and further applied in 
Refs.~\cite{1danQCD,2danQCD,anOPE,CAGC}. This has to do with the fact that
beta function $\beta(a)$ of such IR finite holomorphic coupling $a(Q^2)$
is not fully represented by the
power expansion (\ref{pRGE}), but contains at low $|Q|$ significant nonperturbative
contributions, i.e., contributions nonanalytic in $a$ such as $\exp(-K/a(Q^2))
\sim (\Lambda^2/Q^2)^{K/\beta_0}$. The same is true for the derivative 
on the left-hand side of Eq.~(\ref{RSdepDpt}) when $a_{\rm pt} \mapsto a$.
The equality (\ref{RSdepDpt}) is not valid when we have $a$
(instead of $a_{\rm pt}$) in the theory, the difference between the
left-hand and the right-hand side being a nonperturbative contribution (invisible
to powers of $a$) which tends to get out of control when $N$ is large. 
Hence, additional terms enter the renormalization scale dependence
of the truncated power series in such frameworks and make it even more out of
control at larger $N$. 
On the other hand, it can be shown that the scale dependence for the
reorganized series (mpt) in such frameworks ($a_{\rm pt} \mapsto a$
and $\ta_{\rm pt, n} \mapsto \ta_{n}$)
keeps the simple form (\ref{RSdepDmpt}), i.e., this equality remains
exact in such frameworks. The right-hand side of
Eq.~(\ref{RSdepDmpt}) (with $a_{\rm pt} \mapsto a$) contains the
nonperturbative contributions  - they are contained in the
single term there, the logarithmic derivative $\ta_{N+1}(\kappa Q^2)$
which, in contrast to powers $a(\kappa Q^2)^{N+k}$, ``sees'' such
contributions.

This means that in QCD with the coupling $a(Q^2)$ finite at $Q^2 \to 0$ we
should not use as the basis for the evaluations the power series, but the
reorganized series (man: for ``modified analytic'')
\be
d(Q^2; \kappa)_{\rm man}^{[N]} = a(\kappa Q^2) + 
\sum_{j=1}^{N - 1} \td_j(\kappa) \; \ta_{j+1}(\kappa Q^2) \ ,
\label{DmantrIR}
\ee
where
\be
{\ta}_{n}(Q^2)
\equiv \frac{(-1)^{n-1}}{\beta_0^{n-1} (n-1)!}
\left( \frac{ \partial} {\partial \ln Q^2} \right)^{n-1}  a(Q^2) \ ,
\qquad (n=1,2,\ldots) \ .
\label{tananIR}
\ee
An additional reason for the better convergence and the weaker renormalization 
scale dependence of such series at low $|Q|$ is the empirical fact that in virtually
all models with holomorphic IR finite coupling $a(Q^2)$ we have the
hierarchy $|a(Q^2)| > |\ta_2(Q^2)| > |\ta_3(Q^2)| > \cdots$
for any $Q^2$ (and not just when $|Q^2|$ is large). 

The approach described here was extended in Ref.~\cite{GCAK},
in the frameworks with IR finite holomorphic $a(Q^2)$,
to the evaluation of quantities whose perturbative power expansion (\ref{Dpttr}) 
involves noninteger powers of $a_{\rm pt}$.

It is interesting that in the (F)APT model the evaluation with
the analogs $a^{\rm {(FAPT)}}_{n}(Q^2)$ of the powers of $a_{\rm pt}^{n}$, 
Eqs.~(\ref{MAAnudisp})-(\ref{MAAnu1l}),
is equivalent to the approach described here, because it turns out that
for (F)APT model the relations (\ref{tanan}) are fulfilled
\be
a^{\rm {(FAPT)}}_{n}(Q^2) = {\ta}^{\rm {(FAPT)}}_{{\rm pt},n}(Q^2) 
+ \sum_{m \geq 1} {\tk}_m(n) {\ta}^{\rm {(FAPT)}}_{n+m}(Q^2) \ ,
\label{tananAPT}
\ee
and this even when $n$ is noninteger ($n=\nu$), as argued in Ref.~\cite{GCAK}
(see also Ref.~\cite{KuMagr}, for integer $n$).
Nonetheless, the approach reviewed here can be applied
to general models with holomorphic coupling $a(Q^2)$ with IR finite value,
while the approach Eq.~(\ref{MAAnudisp}) only within the (F)APT.

\section{Numerical evidence}
\label{sec:num}

I will illustrate numerically the effects of various evaluations
in the case of the Adler function in the large-$\beta_0$ approximation.
The effective charge of the (massless) Adler function is defined as
\be
d_{\rm Adl}(Q^2) = - (2 \pi^2) 
\frac{d \Pi(Q^2)}{d \ln Q^2} - 1 \ ,
\label{ddef}
\ee
where $\Pi(Q^2)$ is the correlator of the nonstrange charged hadronic currents
(vector or axial) in the massless limit. The perturbation expansion of $d_{\rm Adl}$
in powers of $a_{\rm pt}$ has the form (\ref{Dpttr}); however, only
the first four coefficients are fully known at the moment 
($d_0=1$; $d_j$ with $j=1,2,3$). I want to test, however, the
renormalization scale dependence and the
convergence of the evaluations based on the truncated perturbation series 
of the type (\ref{Dpttr}) and (\ref{Dmpttr}) [(\ref{DmantrIR})], in pQCD
and in the mentioned IR finite coupling scenarios,  
when the truncation  number $N$ is increasing. The coefficients
$d_n$ and $\td_n$ in $\MSbar$-type schemes can be written as polynomials of
$N_f$ of order $n$, and thus also as polynomials in powers of
$\beta_0$ of order $n$
\be
\td_n(\kappa) = c_{n,n}(\kappa) \beta_0^n + c_{n,n-1} \beta_0^{n-1} + \ldots
+ c_{n,0} \ ,
\label{tdnexpb0}
\ee
The leading-$\beta_0$ (LB) part of these coefficients, 
$\td_{n}^{\rm (LB)} = c_{n,n} \beta_0^n$, are known to all orders
\cite{Broad1,LTM}. This LB quantity can then be written formally
as an integral over momenta $t Q^2$ \cite{Neubert}
\bea
d_{\rm Adl}^{\rm (LB)}(Q^2)_{\rm (m)pt} & = &   
\int_0^{\infty} \frac{dt}{t} \; F_d(t) a_{\rm pt}(t Q^2 e^{{\cal C}})
\label{DLBint}
\\
& = &
a_{\rm pt}(\kappa Q^2) + \td_{1}^{\rm (LB)}(\kappa) {\ta}_{{\rm pt},2}(\kappa Q^2) +
\cdots +  \td_{n}^{\rm (LB)}(\kappa) {\ta}_{{\rm pt}, n+1}(\kappa Q^2) + \cdots
\label{DLBmpt}
\\
& = & a_{\rm pt}(\kappa Q^2) + d_{1}^{\rm (LB)}(\kappa) a_{\rm pt}(\kappa Q^2)^2 + 
\cdots d_{n}^{\rm (LB)}(\kappa) a_{\rm pt}(\kappa Q^2)^{n+1} + \cdots \ ,
\label{DLBpt}
\eea
where $F_d(t)= t {\hat w}_d(t)/4$  is the distribution function of the LB Adler function
obtained in Ref.~\cite{Neubert}, and ${\cal C}=-5/3$ in the
$\Lambda_{\bar MS}$-convention. It is important to point out that
the coupling $a_{\rm pt}$ in the integral (\ref{DLBint}) can
run according to $N$-loop RGE ($N \geq 1$), not just one-loop.
The quantity defined in this way is renormalization scale ($\kappa$) independent,
although it acquires renormalization scheme dependence when
$a_{\rm pt}$ runs according to the $N$-loop RGE with $N \geq 3$ (dependence on
the scheme parameters $c_2, \ldots, c_{N-1}$). Nonetheless, I will
use this quantity for testing the quality of different evaluations,
i.e., evaluations based on the truncated series (\ref{DLBmpt})-(\ref{DLBpt}).
The expansion (\ref{DLBmpt}) is obtained from the integral
representation (\ref{DLBint}) by Taylor-expanding the coupling
$a_{\rm pt}(t Q^2 e^{{\cal C}})$ around the point $\ln \mu^2 \equiv \ln \kappa Q^2$ and exchanging
the order of integration and summation, and using the relations
\bea
{\td}_n^{\rm (LB)}(\kappa) &\equiv &\beta_0^n c_{n,n}(\kappa)
= (\beta_0)^n (-1)^n \int_{t=0}^{\infty} d (\ln t)
\ln^n \left( t \kappa^{-1} e^{{\cal C}} \right) F_d(t) \ , 
\label{tdns1}
\\
c_{n,n}(\kappa)&=&c_{n,n}(e^{\cal C}) + \sum_{k=1}^n 
\left(
\begin{array}{c}
n \\
k
\end{array}
\right)
\ln^k \left( \kappa e^{- {\cal C}} \right) c_{n-k,n-k}(e^{\cal C}) \ ,
\label{tdns2}
\eea
where
\be
c_{n,n}(e^{\cal C}) = \frac{3}{4} C_F \left( \frac{d}{db} \right)^n
P(1-b)|_{b=0} 
\label{cnnAdl}
\ee
with $C_f=4/3$ and $P(x)$ is the trigamma function 
obtained in Ref.~\cite{Broad1}
\be
P(x) = \frac{32}{3 (1 + x)} \sum_{k=1}^{\infty}
\frac{(-1)^k k}{(k^2-x^2)^2} \ .
\label{trigamma}
\ee
While ${\td}_n^{\rm (LB)}$ are the complete LB parts of the full
coefficients  ${\td}_n$, the coefficients
$d_{n}^{\rm (LB)}$ in the power series (\ref{DLBpt}) contain in general
also beyond-the-leading-$\beta_0$ terms. Only in the case of one-loop
RGE running the equality holds: $d_{n}^{\rm (LB)}={\td}_n^{\rm (LB)}$.

In $\MSbar$-type schemes in pQCD the running coupling
$ a_{\rm pt}(t Q^2 e^{{\cal C}})$ in the integral (\ref{DLBint})
has Landau singularities at low $t$, therefore the integral becomes
ambiguous and an integration prescription must be imposed -- usually
the (generalized) principal value which I will adopt here, in order
to define the ``exact'' LB value in pQCD.
On the other hand, in QCD with finite $a(Q^2)$ in the infrared,
all the formulas (\ref{DLBint})-(\ref{DLBpt}) are repeated, with the simple
replacements
\be
a_{\rm pt} \mapsto a \ , \qquad \ta_{{\rm pt},n} \mapsto \ta_{n} \ .
\label{repl}
\ee
Moreover, the exact LB value, i.e., the integral (\ref{DLBint}), 
now becomes finite and unambiguous,
due to the absence of the Landau singularities.

The numerical evaluations will be based on the truncated series
(\ref{DLBmpt}) and (\ref{DLBpt}), for pQCD in the
$c_2=c_3=\cdots=0$ renormalization scheme; and on these truncated series
with the replacements (\ref{repl}) 
for the three frameworks described in Sec.~\ref{sec:3scen}.

\subsection{Stability under the variation of the renormalization scale}
\label{sec:RScl}

\begin{figure}[htb] 
\begin{minipage}[b]{.49\linewidth}
\centering\includegraphics[width=76mm]{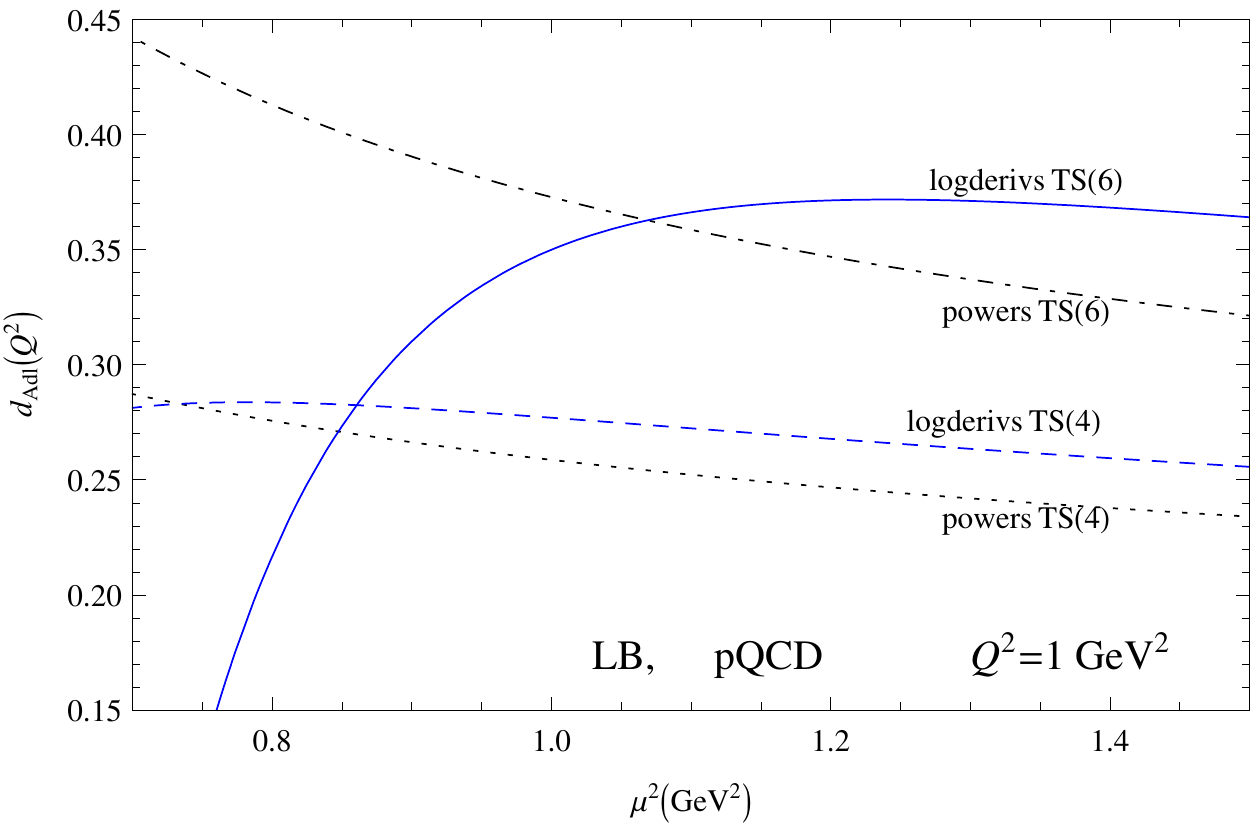}
\end{minipage}
\begin{minipage}[b]{.49\linewidth}
\centering\includegraphics[width=76mm]{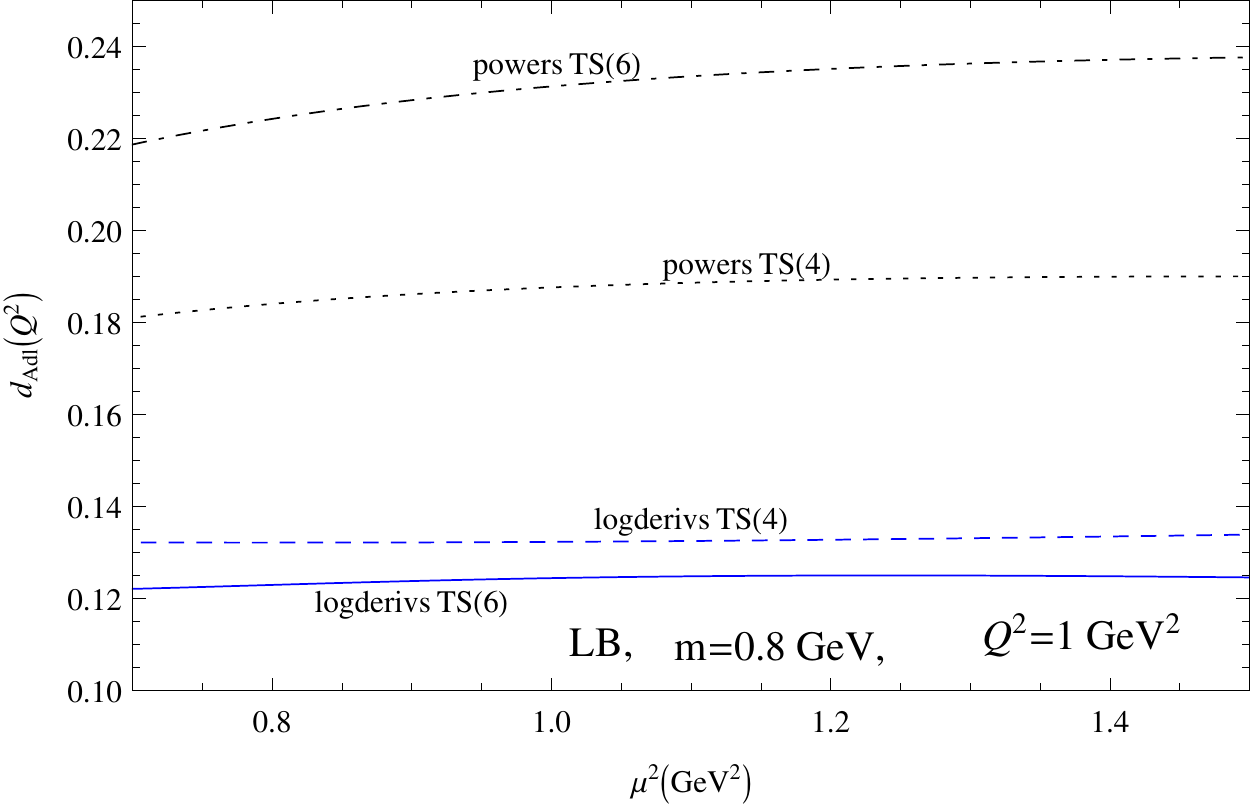}
\end{minipage}
\begin{minipage}[b]{.49\linewidth}
\centering\includegraphics[width=76mm]{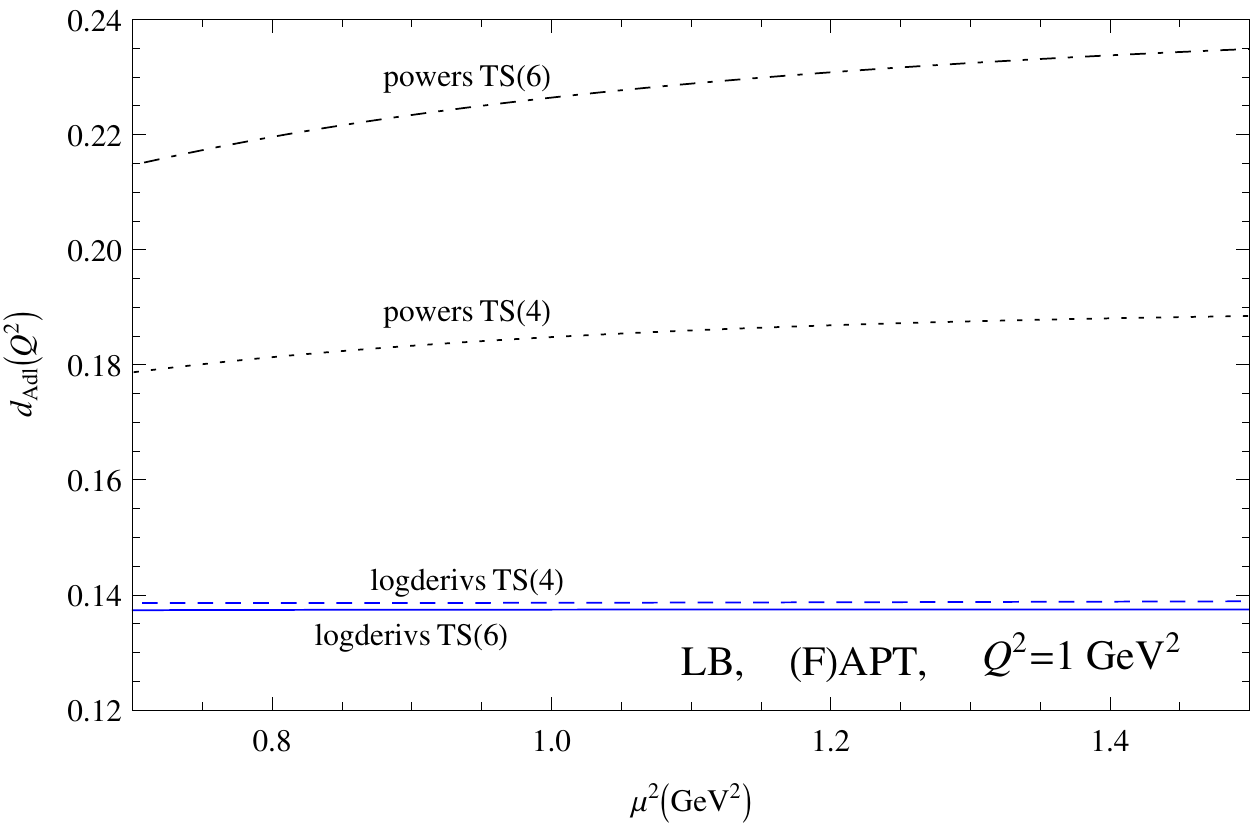}
\end{minipage}
\begin{minipage}[b]{.49\linewidth}
\centering\includegraphics[width=76mm]{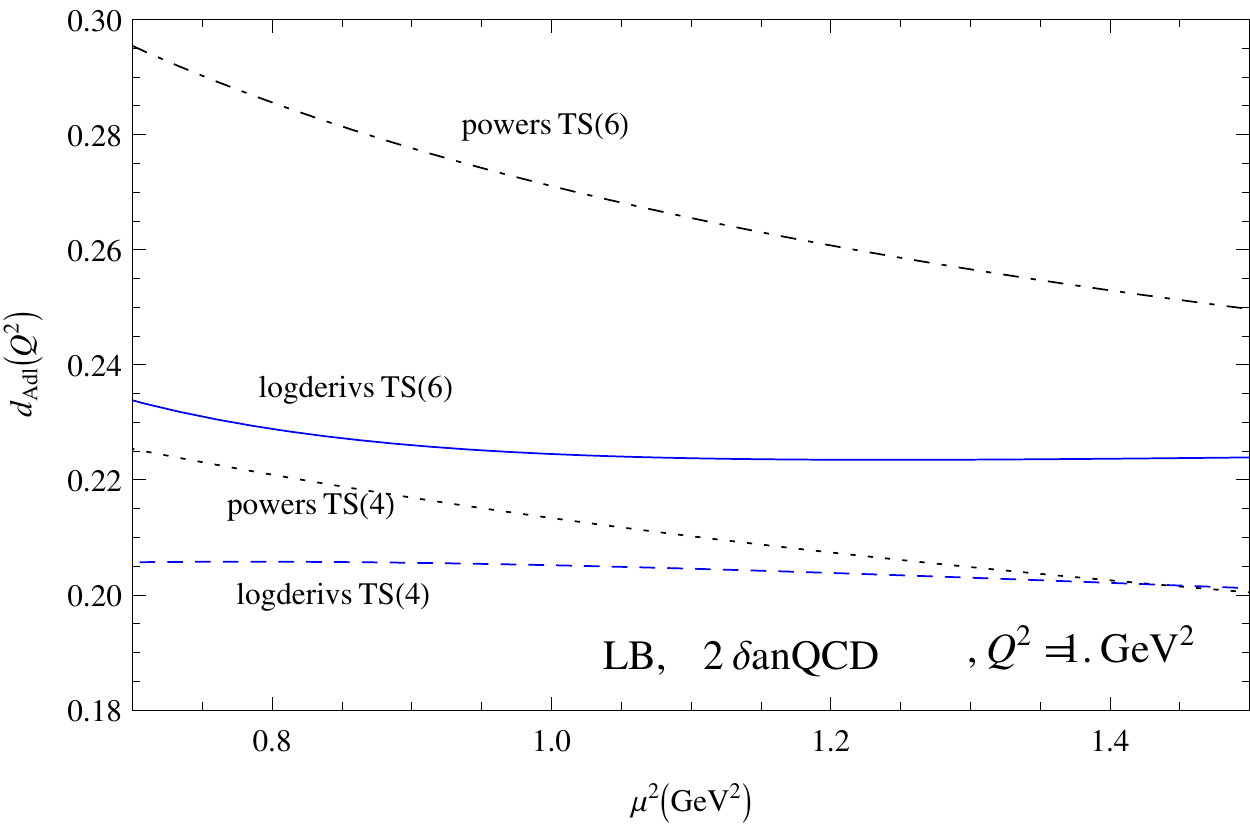}
\end{minipage}
\vspace{-0.2cm}
 \caption{The LB Adler function $d_{\rm Adl}^{\rm (LB)}(Q^2)$,
at leading-$\beta_0$ (LB), for $Q^2=1 \ {\rm GeV}^2$, 
as a function of the (squared) spacelike renormalization scale $\mu^2$:
(a) in pQCD (the upper left-hand Figure);
and in the three frameworks with the coupling $a(Q^2)$ finite in the IR:
(b) the model of Sec.~\ref{sec:dgm} with effective constant gluon mass
 (the upper right-hand Figure);
(c) (F)APT model of Sec.~\ref{sec:fapt} (the lower left-hand Figure);
(d) 2$\delta$ analytic QCD model of Sec.~\ref{sec:2danQCD} (the lower right-hand Figure).
The truncatetions are made at $\sim a^4$ ($\ta_4$) and $\sim a^6$ ($\ta_6$).}
\label{LBvsmu}
 \end{figure}
The results of the LB Adler function, Eqs.~(\ref{DLBmpt})-(\ref{DLBpt})
truncated at order $N=4$ and $N=6$, for $Q^2 = 1 \ {\rm GeV}^2$, are presented
as functions of the squared (spacelike) renormalization scale 
$\mu^2 = \kappa Q^2$ in Figs.~\ref{LBvsmu} for
the pQCD case and
for the three models with IR finite $a(Q^2)$
described in Sec.~\ref{sec:3scen}.
We can see that the truncated series in the logarithmic derivatives
show greater stability under the variation of $\mu^2$ in the
three QCD frameworks with IR finite coupling.

\subsection{Convergence properties of various evaluations}
\label{sec:conv}

Here I will compare the convergence (divergence) behavior of the 
evaluations of truncated series (\ref{DLBmpt})-(\ref{DLBpt})
in pQCD and the three models of  Sec.~\ref{sec:3scen}.
I will add here yet another evaluation method, based on the
truncated series (\ref{DLBmpt}). This method was constructed
in Refs.~\cite{BGApQCD1a,BGApQCD1b} in the context of
pQCD, and was applied with success to QCD frameworks with IR finite holomorphic
$a(Q^2)$ in Refs.~\cite{GCRK,anOPE}. It is an approximation
constructed on the basis of the truncated series in logarithmic
derivatives, truncated at order $\ta_{2 M}$ ($M=1,2,3,\cdots$),
and can be written in the following form:
\be
{\cal G}^{[M/M]}_{d}(Q^2) =  
\sum_{j=1}^M \tal_j \; a(\kappa_j Q^2) \ .
\label{dBG}
\ee
The scale parameters $\kappa_j$ and
the coefficients $\tal_j$ (where: $\tal_1 + \ldots + \tal_M = 1$)
are determined uniquely from the coefficients
$\td_j$ ($j=1,\cdots,2M-1$). I refer for details of the
construction of this expression to the mentioned literature.
Several aspects can be pointed out: (a) the approximant (\ref{dBG}) can be
regarded as a (nontrivial) generalization 
of the diagonal Pad\'e (dPA) method \cite{Gardi}, the latter 
giving renormalization scale independent results at the one-loop level;
(b) the running of $a$ can be to any loop order (not just one-loop),
and the result (\ref{dBG}) is exactly independent of the renormalization
scale used in the original series of logarithmic derivatives; (c) the
approximant fulfills the basic requirement of the 
approximant of order $N=2 M$, Ref.~\cite{GCRK}
\be
d(Q^2) - {\cal G}^{[M/M]}_{d}(Q^2) = {\cal O}(\ta_{2 M+1})
\qquad \left[ = {\cal O}(a^{2 M +1 }) \right] \ .
\label{dBGappr}
\ee
\begin{figure}[htb] 
\begin{minipage}[b]{.49\linewidth}
\centering\includegraphics[width=76mm]{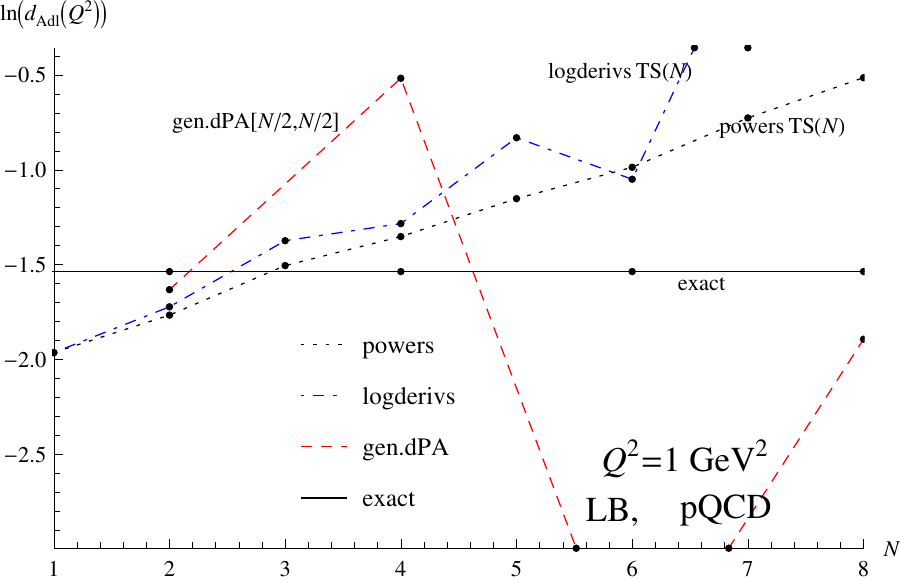}
\end{minipage}
\begin{minipage}[b]{.49\linewidth}
\centering\includegraphics[width=76mm]{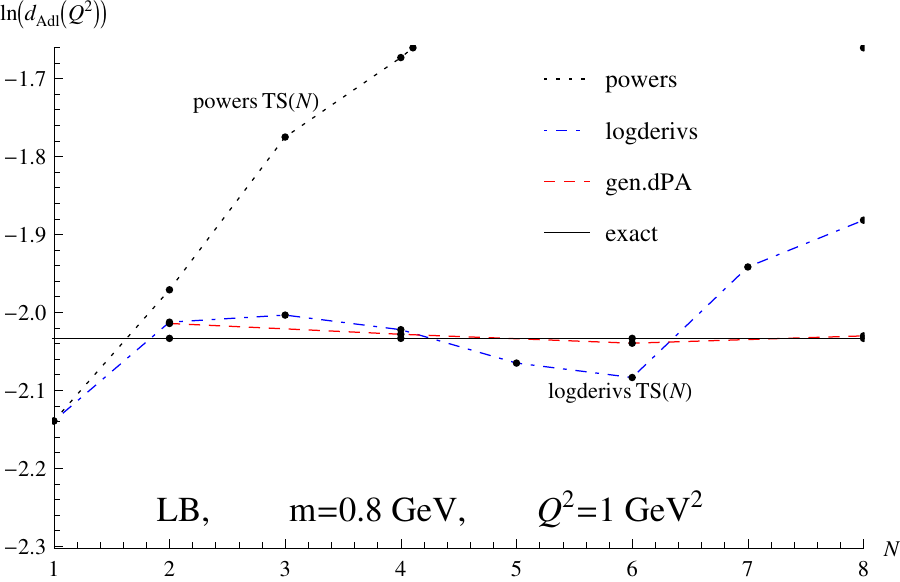}
\end{minipage}
\begin{minipage}[b]{.49\linewidth}
\centering\includegraphics[width=76mm]{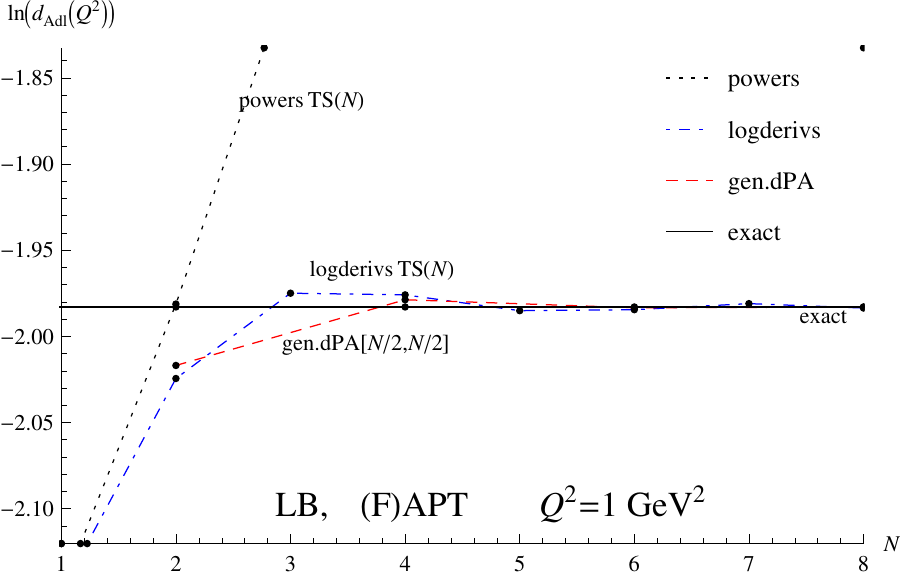}
\end{minipage}
\begin{minipage}[b]{.49\linewidth}
\centering\includegraphics[width=76mm]{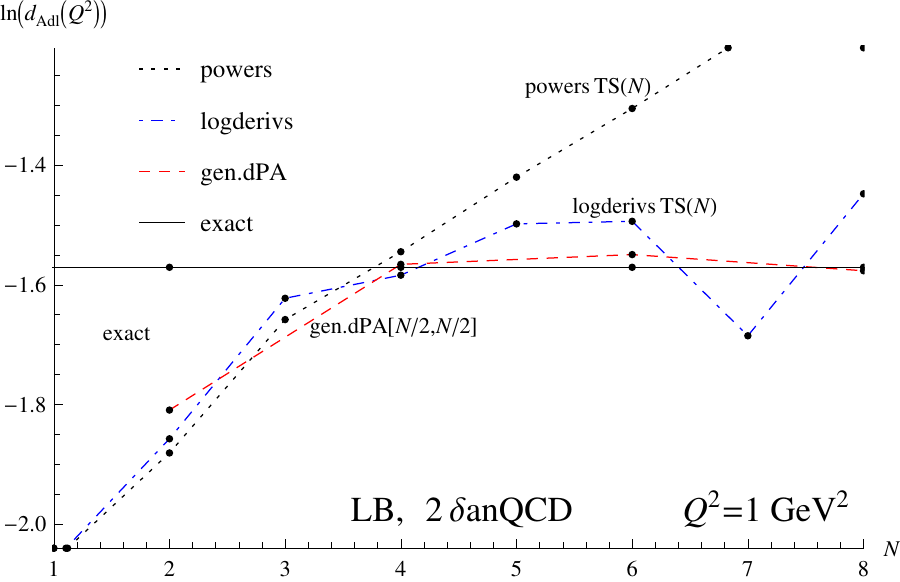}
\end{minipage}
\vspace{-0.2cm}
 \caption{The behavior of the LB Adler function effective charge
$d_{\rm Adl}^{\rm (LB)}(Q^2=1 {\rm GeV}^2)$
as a function of the truncation order $N$
[the vertical axis represents  $\ln d_{\rm Adl}^{\rm (LB)}(Q^2)$]:
(a) pQCD (upper left); 
(b) the model with constant effective gluon mass
of Sec.~\ref{sec:3scen} (upper right); 
(c) (F)APT model of Sec.~\ref{sec:fapt} (lower left);
(d) 2$\delta$ analytic QCD model of Sec.~\ref{sec:2danQCD} (lower right).}
\label{dvLBdBG}
 \end{figure}
I performed the direct evaluations of the
truncated power series (\ref{DLBpt}) and the series in logarithmic
derivatives (\ref{DLBmpt}), as well as the evaluation (\ref{dBG})
[based on the truncated series (\ref{DLBmpt})] for various
orders of truncation $N$, at the chosen renormalization scale
$\mu^2=Q^2$ ($\kappa=1$), in order to see the behavior of these
series with increasing $N$ and to compare the results with
the ``exact'' result (\ref{DLBint}) [with $a_{\rm pt} \mapsto a$ there].
The results are given in Figs.~\ref{dvLBdBG},
for pQCD (in $c_2=c_3 = \cdots = 0$ scheme) and for the three
QCD frameworks of Sec.~\ref{sec:3scen}, at $Q^2=1 \ {\rm GeV}^2$. 
We can see the following: in all three
IR finite frameworks, (a) the naive power series gives highly divergent 
behavior; (b) the series in logarithmic derivatives stabilizes to a degree
at intermediate orders $N \approx 3$-$6$ and then starts to oscillate
increasingly when $N$ increases further;\footnote{
In (F)APT, the series in logarithmic derivatives starts to oscillate
late, at about $N = 10$ which is outside 
the range presented in Figs.~\ref{dvLBdBG}, cf.~Ref.~\cite{GC}.}
(c) the dPA-related method of Eq.~(\ref{dBG})
gives results which converge to the exact LB value (\ref{DLBint}) 
[cf.~also~Eq.~(\ref{repl})] surprisingly well as $N$ increases, 
there is no trace of possible divergent behavior at high $N$ (I checked
this up to $N=20$). 
On the other hand, for the ($\MSbar$-like)
pQCD, all three methods give consistently divergent behavior with increasing
$N$, this being mainly the consequence of the vicinity of the
Landau singularities when $Q^2 \sim 1 \ {\rm GeV}^2$. 
One reason for the failure of the power series (in all cases)
and of the series in logarithmic derivatives (in pQCD already at low $N$;
in IR finite framework at high $N > 10$) is the renormalon growth of
the coefficients $c_{n,n} \sim n!$. The dPA-related method (\ref{dBG}),
on the other hand, appears to deal with the renormalon growth of
the coefficients very well, and the only problem for that method are
the Landau singularities which, in the frameworks with holomorphic
(analytic) and IR finite $a(Q^2)$ are nonexistent. Even more, this
dPA-related method, which is based on the truncated series in 
logarithmic derivatives, is completely renormalization scale independent,
i.e., in Figs.~\ref{LBvsmu} it would be represented by exactly horizontal lines.

\section{Timelike observables}
\label{sec:time}

The timelike observables ${\cal T}(\sigma)$, 
such as cross sections and decay widths,
can be related with spacelike observables
${\cal F}(Q^2)$, via integral transformations.

Often the integral transformations between ${\cal T}(\sigma)$ and
${\cal F}(Q^2)$ are the same or similar as
between the $(e^+ e^- \to \ {\rm hadrons})$ ratio 
${\cal T}(\sigma)=R(\sigma)$ and the
Adler function ${\cal F}(Q^2)=d_{\rm Adl}(Q^2)$ 
\begin{equation}
{\cal F}(Q^2) = Q^2 \int_0^{\infty} 
\frac{d \sigma \ {\cal T}(\sigma)}{(\sigma + Q^2)^2} \ ,
\qquad 
{\cal T}(\sigma) = \frac{1}{2 \pi i} 
\int_{-\sigma - i \varepsilon}^{-\sigma + i \varepsilon} 
\frac{d Q^{' 2}}{Q^{' 2}} {\cal F}(Q^{' 2}) \ ,
\label{FTTF}
\end{equation}
where in the last integral the integration contour is in the complex 
$Q^{' 2}$-plane encircling the singularities of the integrand;
for example, on the circle of radius $\sigma$ in the
counterclockwise direction (and not cutting the negavive semiaxis).

The basic idea for evaluations of such timelike quantities in
the QCD frameworks with analytic and IR finite $a(Q^2)$ is
that first the spacelike quantity ${\cal F}(Q^{' 2})$ is evaluated
(with $Q^{' 2}$ on the mentioned circle), with aforementioned method
of truncated series in logarithmic derivatives, or the dPA-related
method (\ref{dBG}); then the contour integral (\ref{FTTF}) is
applied on this quantity.

\section{Summary}
\label{sec:summ}

Theoretical approaches such as Dyson-Schwinger equations and other
functional methods, most of the analytic (holomorphic) QCD models,
as well as lattice calculations, suggest that the QCD running coupling 
$a(Q^2)$ ($\equiv \alpha_s(Q^2)/\pi$) is finite in the IR limit $Q^2 \to 0$.
Here, it was argued that in such frameworks the evaluation of the
renormalization scale invariant spacelike QCD quantities $d(Q^2)$,
at low $|Q^2| \sim 1 \ {\rm GeV}^2$,
should not be performed as a naive truncated power series, but rather
as a truncated series in logarithmic derivatives, 
cf.~Eqs.~(\ref{DmantrIR})-(\ref{tananIR}). The reason for this lies in
the fact that the powers do not take into account correctly the 
nonperturbative (nonanalytic in $a$) terms, and this is reflected in
the increasingly strong renormalization scale dependence when the number of
power terms increases. The logarithmic derivatives, on the other hand,
take into account the nonperturbative terms in a systematic way,
and the scale dependence of such truncated series does not increase 
due to such terms (which are under control in this case) 
but only due to the renormalon 
growth of the coefficients. 
Further, in such frameworks, the evaluation method of Eq.~(\ref{dBG}), which is
based on the truncated series in logarithmic derivatives and can be regarded as 
a generalization of the diagonal Pad\'e method, gives results which are exactly 
renormalization scale independent and show very good convergence properties as 
the number of terms increases. Numerical evidence for all these arguments was 
presented for the leading-$\beta_0$ (LB) Adler function $d_{\rm Adl}^{\rm (LB)}(Q^2)$, 
which is a renormalization scale invariant quantity in all such frameworks.

It is, however, realistic to assume that such QCD frameworks, with 
finite $a(Q^2)$ when $Q^2 \to 0$, do not give us all the nonperturbative
effects in the ``perturbative'' leading-twist term, and that other
nonperturbative contributions should be added, either via higher-twist
terms of OPE \cite{Shifman:1978bx}, or by directly including
such contributions in the specific considered observables 
\cite{DeRafael,MagrDual,Milton:2001mq,mes2,Nest3a,Nest3b}
(see also: \cite{Deuretal,Court1,Court2}). If applying
OPE in QCD with holomorphic and IR finite coupling $a(Q^2)$,
it is preferable that $a(Q^2)$ differs very little from the underlying
($\MSbar$-like) perturbative coupling $a_{\rm pt}(Q^2)$ at high $|Q^2|$,
in order to maintain the ITEP School interpretation \cite{Shifman:1978bx}
of the OPE higher-twist terms as being exclusively of
the IR origin. In Ref.~\cite{2danQCD} we constructed such a model
in which $a(Q^2)- a_{\rm pt}(Q^2) \sim (\Lambda^2/Q^2)^5$ at large $|Q^2|$,
and applied it with OPE in Ref.~\cite{anOPE}.

\begin{acknowledgements}
This work was supported in part by FONDECYT (Chile) Grant No.~1130599.
\end{acknowledgements}



\end{document}